\documentclass[12pt,a4paper]{article}

\usepackage{a4}
\usepackage{amsfonts}
\usepackage{amssymb}
\usepackage{epsfig}
\usepackage{psfig}
\usepackage{psfrag}

\newcommand{\beqn}{\begin{eqnarray}}
\newcommand{\eeqn}{\end{eqnarray}}
\newcommand{\be}{\begin{equation}}
\newcommand{\ee}{\end{equation}}
\newcommand{\non}{\nonumber \\}
\newcommand{\om}{\overline{m}}
\begin{document} 

\title{}
\begin{flushright}
\vspace{-3cm}
{\small HU-EP-02/05 \\
        SPIN-02/03 \\
        ITP-UU-02/01 \\ 
        hep-th/0202024}
\end{flushright}
\vspace{1cm}

\begin{center}
{\Large\bf Moduli Stabilization for Intersecting Brane \\[.3cm]
          Worlds in Type 0$'$ String Theory }
\end{center}

\vspace{0.5cm}

\author{} 
\date{}
\thispagestyle{empty}

\begin{center}
{\bf Ralph Blumenhagen},\footnote{e-mail: blumenha@physik.hu-berlin.de}$^{,*}$ 
{\bf Boris K\"ors}\footnote{e-mail: kors@phys.uu.nl}$^{,\dag}$
{\bf and Dieter L\"ust}\footnote{e-mail: luest@physik.hu-berlin.de}$^{,*}$
\\
\vspace{0.5cm}
{\it 
$^*$Humboldt Universit\"at zu Berlin \\
{\small{Institut f\"ur Physik, Invalidenstr. 110, 10115 Berlin, Germany}} \\
\vspace{.3cm} 
$^\dag$Spinoza Institute, Utrecht University \\
{\small{Utrecht, The Netherlands}} 
}
\vspace{1cm}
\end{center}                  

\begin{center}
{\bf Abstract} \\
\end{center}
Starting from the non-supersymmetric, tachyon-free  orientifold 
of type 0 string theory, we construct four-dimensional brane world models
with D6-branes intersecting at angles on internal tori.
They support phenomenologically interesting gauge theories with chiral
fermions.
Despite the theory being non-supersymmetric the perturbative scalar potential
induced
at leading order is shown to stabilize  geometric moduli, leaving only the
dilaton tadpole uncanceled.
As an example we present a three generation model with gauge
group and fermion spectrum close to a left-right symmetric
extension of the Standard Model.

\clearpage

\section{Introduction} 

Since it was 
discovered that one can have string excitations as light as a 
few TeV by allowing for large internal compactification spaces 
\cite{hepph9803315,hepph9804398}, 
string models without supersymmetry received new attention as 
phenomenologically plausible scenarios. 
This may either involve string theories without any supersymmetry already in 
ten dimensions or compactifications of originally supersymmetric 
theories with supersymmetry breaking at the string scale. 
A crucial problem in these constructions is the dynamical stability of the 
background which is no longer protected by any kind of no-force law. 
The most radical signal of an inherent instability is the presence of 
tachyons in the spectrum, scalars 
with a negative tree level mass squared. By now, the open string tachyons 
localized 
on some lower dimensional defect, a D-brane, in the entire ten-dimensional 
space-time are rather well understood 
\cite{hepth9805170,hepth9911116,hepth9912249}. 
They are unstable modes of the gauge theory sector and 
their condensation translates into a decay of these D-branes 
into stable configurations but does not 
affect equally drastically the background space which the branes are 
wrapped on. 
The latter  is indeed the case for closed string tachyons, unstable 
gravitational modes, 
which directly relate to a decay of the space-time. 
In  field theoretic language, open string tachyon condensation and the
induced D-brane recombination has its natural interpretation
as the Higgs effect, namely the spontaneous breaking of the corresponding
gauge group to that subgroup, which is associated to the stable D-brane
configuration at the endpoint of the decay.\footnote{See also 
\cite{hepth0012157,hepth0107036} 
in this context.} Hence open string tachyons may indeed
occur but are not harmful as they can play the role of the
Higgs bosons of the effective gauge theory. \\

While one can often read the statement that a non-supersymmetric 
configuration 
is already considered stable if tachyons are absent, this is of course 
still 
a very simplistic view. In the absence of supersymmetry, quantum corrections 
will 
usually tend to generate potentials for the other
massless scalar fields that very often 
drive 
the background either to some extreme limit or to a point where tachyons
reappear. In particular the dilaton can be pushed to 
zero or infinite coupling. 
It will be one of the main objectives of this paper to construct 
non-supersymmetric D-brane models in which the geometrical moduli
fields are stabilized.\\

There exist two possibilities to construct non-supersymmetric
brane world models in the first place: $(i)$ One starts from a 
supersymmetric theory in ten dimensions and only breaks supersymmetry 
in the open string, gauge 
theory sector on the branes. $(ii)$ Supersymmetry is already broken
right from the very beginning in the closed string sector.
Most of the previous work follows the first pattern,
namely one is considering supersymmetric type I and type II string theory
where D6-branes intersect at relative angles on an internal torus or orbifold,
which has been first proposed 
in \cite{hepth0007024,hepth0007090} and developed further in 
\cite{hepth0010279,hepth0011073,hepph0011132,hepth0012156,
hepth0105155,hepth0107138,hepth0107166,hepth0108131,hepph0109082,hepth0112015,
hepth0201037,hepth0201205}. 
These models display many attractive, generic features such as 
chiral fermion spectra in the four-dimensional effective theory 
\cite{hepth9503030,hepth9606139}, which are localized at intersections
of two D6-branes and therefore mostly transform in the  
bifundamental representation of the  unitary gauge groups living on the 
respective D6-branes. Scalar fields either decouple or become tachyonic and 
may serve as Higgs bosons, as their condensation closely resembles standard 
spontaneous symmetry breaking patterns. In this way intersecting D-brane
scenarios can be constructed which come very close to the
non-supersymmetric Standard Model. However 
in the simplest case of D6-branes wrapped around homology 3-cycles
of the six-dimensional torus, the background geometry is generically
unstable. \\

In this paper we will discuss the second option $(ii)$
as an alternative, novel approach for intersecting brane world models.
To be concrete we will construct a brane world scenario of type 0$'$ string
theory
\cite{hepth9509080,hepth9701137,hepth9702093,hepth9810214,hepth9904069,
hepth9906234,hepth9909010} 
with intersecting D6-branes, where we will closely follow
the type I and type II D-brane  models  described before. The
low energy description of our type 0$'$ brane worlds has many appealing
features 
close to Standard Model or GUT physics and at the same time stabilizes at least
some of the closed string geometric moduli.
In fact, while one may expect that these type 0$'$
compactifications could lead to models with 
worse stability properties compared to the type II and
type I D-brane models, the perturbative analysis  
appears to suggest the opposite: The formerly constructed toroidal 
intersecting 
brane worlds of type I string theory suffer from a perturbative instability 
that drives 
the complex structure moduli $U_2^I$ of the torus to a degenerate limit, 
$U_2^I \rightarrow \infty$, 
which implies an unacceptable squashing of the internal space. 
This instability could only be avoided by freezing these fields to particular 
values 
by imposing orbifold symmetries \cite{hepth0107138,hepth0112015}. 
We shall find that the type 0$'$ brane worlds 
do not display this problem, but, on the contrary, their scalar potential 
stabilizes 
the $U_2^I$ at finite specific  values, leaving only the dilaton 
tadpole uncanceled 
at leading order. At the same time they share all the generic properties of 
the 
former intersecting brane worlds of type I strings and only introduce slight 
modifications in the explicit construction. Of course, all of these statements 
can be subject to higher order perturbative or even non-perturbative 
corrections, which 
are hoped to stabilize the dilaton somehow, but may also affect the other 
moduli. 
Finally, open string tachyon condensation may again occur and play the
role of the Higgs effect in the spontaneously broken gauge sector.
However now the D-brane configuration after tachyon condensation
will still be non-supersymmetric.
\\

The paper is organized as follows: In section \ref{ch2} we briefly review the 
construction of the ten-dimensional tachyon-free type 0$'$ string theory 
and its basic properties. We then discuss the modifications that come into 
play by 
employing the intersecting brane world concept within this theory in section 
\ref{ch3}. 
Next we present the spectra of massless fermions, study the requirement of 
anomaly 
cancellation as a consistency check and discuss the scalar potential of the 
complex structure moduli in section \ref{ch4}. Finally, in section 
\ref{ch5}, we we   
present an example with the gauge group of a left-right symmetric extension  
of the Standard Model and corresponding spectrum of chiral fermions, whose 
complex structure moduli are being frozen at specific  values.

\section{Type 0$'$ string theory} 
\label{ch2}

The starting point is the non-supersymmetric ten-dimensional type 0B string
theory \cite{DH1986,SW1986},  
which is known to be perturbatively unstable due its tachyonic ground state.  

\subsection{Type 0B string theory}

Type 0 string theory can either be defined as a quotient  of
type II string theory by the space-time fermion number $(-1)^{F_s}$.
Alternatively, one can start with the ten-dimensional superstring theory 
equipped with the modified GSO projection
\beqn
P_{\rm GSO} = \frac{ 1 + (-1)^{F_L +F_R} }{2}
\eeqn
leading to the type 0B torus amplitude
\beqn
{\cal T} &\sim& {1\over \vert \eta \vert^{16}}\, \left( 
\vert O_8 \vert^2 + \vert V_8 \vert^2 + 
\vert S_8 \vert^2 + \vert C_2 \vert^2 \right) \non
&\sim& {1\over 2\,\vert \eta \vert^{24}}\, \left(
\vert {\vartheta[ {\textstyle{0 \atop 0}} ] } \vert^8 +
\vert {\vartheta[ {\textstyle{0 \atop 1/2}} ] } \vert^8+
\vert {\vartheta[ {\textstyle{1/2 \atop 0}} ] } \vert^8 \right).
\eeqn 
We have denoted by  $O_8$ etc. the characters of the $SO(8)$ affine
Lie-Algebra at level $k=1$.
One peculiarity of this model is that compared to type II superstring theory
all Ramond-Ramond (RR) forms are doubled. 
Type 0B string theory therefore contains even RR-forms $C^\pm$ originating
from the (R$+$,R$+$) respectively (R$-$,R$-$) sector, where the sign indicates
the left respectively right moving world-sheet fermion number. 
Since each RR-form appears twice, there exist also two kinds of 
D$(p-1)$-branes,
which couple to the fields $C^\pm_p$ respectively.  
These D-branes are most conveniently be  described in another basis, namely
\beqn
             C_p={1\over \sqrt 2}\left( C_p^+ + C_p^- \right),\quad\quad
             C'_p={1\over \sqrt 2}\left( C_p^+ - C_p^- \right),
\eeqn
where the two D$(p-1)$-branes are given by the boundary states 
\cite{PC1987,CLNY1988,hepth0005029}
\beqn 
\vert {\rm D}(p-1) , \eta,\eta' \rangle = \vert {\rm D}(p-1) ,
\eta \rangle_{\rm NSNS} 
+ \vert {\rm D}(p-1) ,\eta' \rangle_{\rm RR} . 
\eeqn
with $\eta=\eta'=1$ for a D$(p-1)$-brane and $\eta=\eta'=-1$ for a
D$(p-1)'$-brane. 
The freedom of choice of the  sign $\eta$ is due to  
the boundary condition of a world sheet fermion at the position of 
a brane
\beqn 
\psi^i_r - i\eta \tilde\psi_r^i =0 ,\quad \psi^\mu_r + i\eta 
\tilde\psi_r^\mu =0 , 
\eeqn
with indices $i$ referring to directions transverse to the brane and 
indices $\mu$ 
along the brane. Note, that in type II string theory only the superposition 
D$(p-1)$+D$(p-1)'$ is invariant under the GSO projection, so that only one
D-brane of each even dimensionality survives. 

\subsection{Type 0$'$ string theory}

In \cite{hepth9509080,hepth9702093} it was noted for the first
time that by taking a particular orientifold of type 0 string theory
one can get rid of the closed string tachyon.
Dividing by the world-sheet parity $\Omega$ alone does not remove the tachyon
but the combination $\Omega' = \Omega (-1)^{F_R}$ 
\cite{hepth9904069} does. 
Let us review what the effect of this projection on the RR-forms and D-branes
is. The RR-forms $C_p^\pm$ transform
as
\beqn \label{act}
\Omega:&\ &C_p^\pm \to (-1)^{(p-2)\over 2}\, C_p^\pm, \non
(-1)^{F_R}:&\ & C_p^\pm \to \pm C_p^\pm,
\eeqn
so that the combined action gives
\beqn \label{actb}
\Omega':&\ &C_p^\pm \to \pm (-1)^{(p-2)\over 2}\, C_p^\pm.
\eeqn
Thus, the following forms survive the orientifold projection
\beqn \label{actc}
C_{10}^+, C_{8}^-,C_{6}^+,C_{4}^-,C_{2}^+,C_{0}^- . 
\eeqn
Summarizing, in type 0$'$ string theory there exist  only one RR-form
of each even degree. The action of $\Omega'$  on the D-branes is
\beqn
\Omega' \vert {\rm D}p, \eta,\eta' \rangle = 
\vert {\rm D}p,-\eta,-\eta'\rangle 
\eeqn
implying that the symmetry of the brane spectrum under $\Omega'$ 
requires any respective D$p$-brane to be accompanied by a D$p'$-brane. \\

Using these inputs, in \cite{hepth9509080,hepth9904069} 
the complete string amplitudes at
one loop order have been 
evaluated and the tadpole cancellation conditions deduced. 
They demand the introduction of $N + N' =64$ 
branes, i.e. 32 of each type, and imply a maximal gauge symmetry $U(32)$. 
In the type I superstring theory $\Omega$ leaves the branes invariant and 
thus imposes a projection upon the open string 
excitations that leads to a gauge group $SO(32)$ . 
Here, $\Omega'$ permutes the two kinds of branes and identifies their  degrees 
of freedom, so that one is left with a unitary gauge group $U(32)$. \\
 
A property of this theory which will become important later on 
is that the orientifold
planes do only couple to closed string modes from the RR-sector, implying that 
they have vanishing tension. This can be seen from the tree-channel
Klein-bottle and M\"obius strip amplitudes
\beqn \label{kb}
{\widetilde{\cal K}} &=& - 2^{10}\, c \int_0^\infty{dl\ 
\frac{\vartheta\big[ {1/2 \atop 0} \big]^4}{\eta^{12}} } , \\ 
\label{ms}
{\widetilde{\cal M}} &=&  2^5 \left( N + N' \right) c \int_0^\infty{dl\
\frac{\vartheta\big[ {1/2 \atop 0} \big]^4}{\eta^{12}} } , 
\eeqn
with arguments $\tilde q=\exp(-4\pi l)$ and $\tilde q=-\exp ( -4\pi l)$ 
respectively 
and the normalization $c = {\rm Vol}_{10} / \left( 8\pi^2 \alpha'\right)^5$. 
This leads to a modification of the scalar potential for the 
moduli fields in the way that the orientifold tension of type I 
string theory does not appear there. \\   

The interaction between the D-branes at leading order are  given
by the annulus amplitudes. In the loop channel the amplitude between
two D-branes of the same kind looks like
\beqn \label{an}
{\cal A}_{({\rm D}p,{\rm D}p)} 
= \frac{N^2 + {N'}^2}{8} c \int_0^\infty{\frac{dt}{t^6}{
\frac{\vartheta\big[ {0 \atop 0} \big]^4 - 
\vartheta\big[ {0 \atop 1/2} \big]^4}{
\eta^{12}} }} 
\eeqn
with argument $q =\exp(-2\pi t)$.
This  leads to a repulsive force between the two D-branes, which means that
the tension of the type 0 branes is not any longer balanced against
their RR-charge. Moreover, it is evident from (\ref{an}) that 
between two D-branes of the same type there are only
open string excitations which are bosonic in space-time.
Contrary, for the loop channel annulus amplitude for two D-branes of
opposite type one obtains
\beqn \label{anb}
{\cal A}_{({\rm D}p,{\rm D}p')} 
= - \frac{N N'}{4} c \int_0^\infty{\frac{dt}{t^6}{
\frac{\vartheta\big[ {1/2 \atop 0} \big]^4}{\eta^{12}} }}
\eeqn
leading to space-time fermions. Note, that even though the closed
string sector is purely bosonic the open string sector
contains space-time fermions. Moreover, the force between two D-branes of
opposite type is attractive, as they only interact via exchange of closed
string modes from the NSNS sector. This is clear, since the two D-brane are
not charged under a common RR-form. However, adding the two annulus amplitudes
yields a vanishing force similar to the no-force BPS situation, 
from which we can deduce that the tension of the type 0 branes
is related to the tension of the type II branes via
\beqn \label{ten}
 T_0={T_{\rm II}\over \sqrt 2}.
\eeqn

The RR-tadpole cancellation condition is satisfied for $N=N'=32$ 
leading to a gauge group $U(32)$ with additional massless
Majorana-Weyl fermions in the ${\bf 496} \oplus \overline{\bf 496}$ 
representation of $U(32)$. Note, that the annulus is also the only amplitude 
which receives contributions from the NSNS sector in the tree channel.  
In ten dimensions this is attributed to the dilaton tadpole exclusively, 
which leads to a run-away behavior for the string coupling. 
An effective way to deal with it may 
consist in a suitable adaption of the Fischler-Susskind mechanism 
\cite{FS1986,FS1986b,hepth0004165,hepth0011269}. 
In lower dimensions the scalar potential will also involve 
geometric moduli fields. 

\section{Type 0$'$ intersecting brane worlds}
\label{ch3}

In \cite{hepth0007024} magnetic background fields on internal tori were shown 
to provide effective means to construct interesting type I string 
compactifications 
with chiral fermions in unitary gauge groups.
By an exact perturbative duality transformation, a T-duality along three of 
the internal circles, this setting of (non-commutative) D9-branes with 
magnetic 
background fluxes is transformed into a completely equivalent picture with 
(commutative) 
D6-branes intersecting at relative angles on the dual tori 
\cite{hepth0003024,hepth0010198}.    
The later investigations of the prospects and properties of these models 
revealed 
them to be among the most promising candidates to achieve a bottom-up 
construction of a Standard Model or GUT field theory out of string theory. 
While most effort was spent on studying non-supersymmetric models, 
some attention was also devoted to supersymmetric constructions 
\cite{hepth0107143,hepth0107166,hepth0111179} 
which have 
close relations to M-theory vacua with a background space of $G_2$ 
holonomy. \\ 

In this fashion we now consider a compactification on an internal 
six-dimensional 
complex torus 
\beqn 
\mathbb{T}^6 = \mathbb{T}^2_1 \times \mathbb{T}^2_2 \times \mathbb{T}^2_3 
\eeqn
with coordinates $(X^I,Y^I)$ on each $\mathbb{T}_I^2$. 
Each two-dimensional 
torus is defined by its complex structure $U^I = U_1^I + i U_2^I$ and its 
K\"ahler structure $T^I = T_1^I + i T^I_2$. 
In order to apply the methods of intersecting brane worlds to the type 0$'$ 
theory, 
one also needs to switch to a T-dual theory. The world sheet parity projection 
gets mapped to $\Omega' {\cal R}$ where ${\cal R} : Y^I \mapsto -Y^I$ is a 
reflection 
along the three circles that have been dualized. 
The theory with maximal gauge group now simply contains 32 D6- and 
32 D6$'$-branes 
located along the fixed circles of ${\cal R}$, i.e. along $X^I$, to cancel the
tadpoles.
Note, that the discrete parameters $U_1^I=b^I$ can take two values
$b^I\in\{0,1/2\}$ in order to maintain the ${\cal R}$ symmetry of 
the background, corresponding to the two tori shown in figure 1. \\ 

\begin{figure}[h] 
\begin{center}
\psfrag{x1}[bc][bc][.8][0]{$X^I$} 
\psfrag{x2}[bc][bc][.8][0]{$Y^I$} 
\psfrag{A}[bc][bc][.8][0]{$\quad\quad b^I=0$} 
\psfrag{B}[bc][bc][.8][0]{$\quad\quad b^I=\frac{1}{2}$} 
\makebox[10cm]{
 \epsfxsize=10cm
 \epsfysize=5cm
\epsfbox{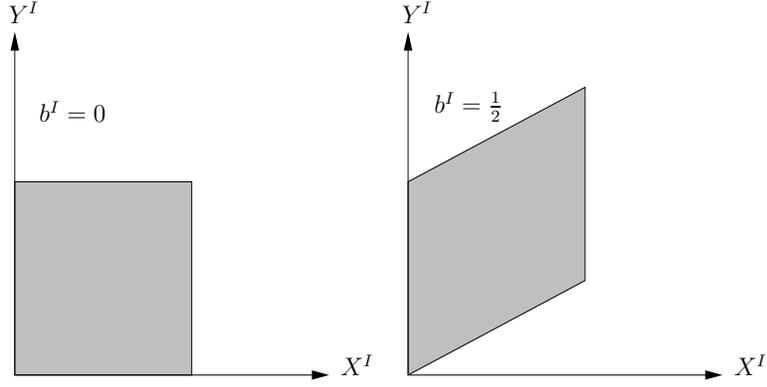}
}
\end{center}
\caption{Torus with $b^I=0,1/2$}
\label{figz2latt}
\end{figure}
%

The most general D6-brane of type $a$ 
wrapped on the $\mathbb{T}^6$ and of codimension one 
on any single $\mathbb{T}^2_I$ is characterized by its winding numbers 
$(n_a^I,m_a^I)$ on the six elementary circles. It is appropriate 
to use a basis 
\beqn 
(n_a^I,\overline{m}_a^I)=(n_a^I,m_a^I+b^I\,n_a^I)
\eeqn
denoting the wrapping around the  cycles along  $X^I$ and $Y^I$
respectively.  
The parity operation $\Omega'{\cal R}$ acts also on the winding numbers  
\beqn \label{sym}
\Omega' {\cal R}:\ {\rm D}6_{(n_a^I,\overline{m}_a^I)} \mapsto 
 {\rm D}6_{(n_a^I,-\overline{m}_a^I)} . 
\eeqn
This is illustrated in figure 2. \\

\begin{figure}[h] 
\begin{center}
\psfrag{x1}[bc][bc][.8][0]{$X^I$} 
\psfrag{x2}[bc][bc][.8][0]{$Y^I$} 
\psfrag{a}[bc][bc][.8][0]{$a$} 
\psfrag{b}[bc][bc][.8][0]{$b$} 
\psfrag{ap}[bc][bc][.8][0]{$a'$} 
\psfrag{bp}[bc][bc][.8][0]{$b'$} 
\makebox[10cm]{
 \epsfxsize=10cm
 \epsfysize=5cm
\epsfbox{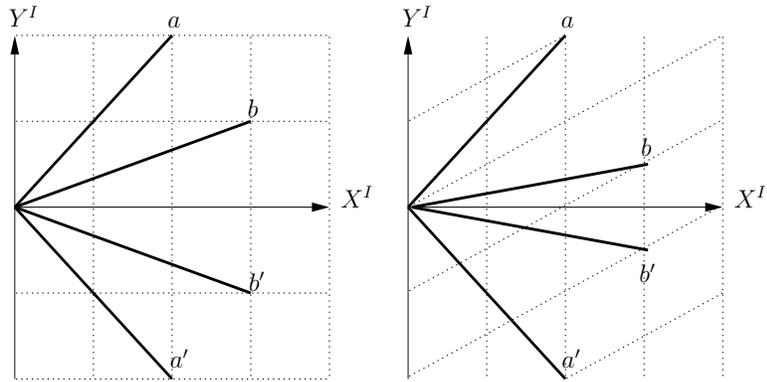}
}
\end{center}
\caption{D6-brane configurations}
\end{figure}
%

Thus the symmetry of the brane spectrum  requires to combine any D6-brane 
at some 
angle 
\beqn
\varphi^I_a = {\rm arctan}
\left( \frac{\overline{m}_a^I}{n_a^I U_2^I} \right)
\eeqn
relative to the $X^I$ axis with another D6$'$-brane at an opposite angle 
$-\varphi_a^I$. 
We henceforth employ the convention to use letters $a=1...K$ for the branes of 
type $\eta=\eta'=1$ and $a'=1...K$ for those of type $\eta=\eta'=-1$.  
Since the two sets of 
stacks are identified under $\Omega'{\cal R}$,  there will be a gauge group 
$U(N_a)$ on any individual stack with $N_a = N_{a'}$ being the number of 
branes. Similar to  the type I models we expect 
chiral fermions at any intersection point of such branes. \\ 

In order to get a precise 
quantitative understanding we need to embark on the computation of the  
contributions to the tadpole divergencies at one loop order in string 
perturbation  theory. 
The three relevant diagrams can most easily be found by combining the 
results of \cite{hepth9904069} and \cite{hepth0007024} and have been collected 
in the appendix. 
The upshot of this computation  is summarized in the following: 
The Klein bottle does not involve any open strings. The oscillator part is 
thus 
given by (\ref{kb}) but momentum integrations need to be replaced by 
Kaluza-Klein (KK) 
and winding sums. One then gets
\beqn
{\widetilde{\cal K}} = 2^{11}\, c_4\, \prod_{I=1}^3{\left( U_2^I \right)} 
\int_0^\infty{dl}\ +\ {\rm finite} 
\eeqn
for the tree channel contribution to the RR tadpole. 
We have  defined $c_4 = {\rm Vol}_4 / \left( 8\pi^2 \alpha' \right)^2$. 
Note again that there is no NSNS contribution whatsoever. 
The open string amplitudes involve $({\rm D}6,{\rm D}6)$ 
and $({\rm D}6,{\rm D}6')$ open strings 
separately. The M\"obius strip receives only contributions from open strings 
invariant under $\Omega' {\cal R}$, i.e. those stretching between branes of
opposite type. Its RR tadpole reads
\beqn
{\widetilde{\cal M}}_{aa'} = -2^6\, c_4\, N_{a} 
                   \prod_{I=1}^3{\left( n_a^I U_2^I\right)} 
\int_0^\infty{dl}\ +\ {\rm finite}. 
\eeqn
Again, there is no NSNS tadpole. 
Finally, the annulus diagram involves all kinds of open strings, where the
contribution to  the RR tadpole arises merely from the $({\rm D}6,{\rm D}6)$ 
and $({\rm D}6',{\rm D}6')$
open strings 
\beqn
{\widetilde{\cal A}}^{({\rm RR})}_{ab} = 
 c_4\, N_a\, N_b\,  \prod_{I=1}^3 \left( n_a^I\, n_b^I\,
    U_2^I+\overline{m}_a^I\, \overline{m}_b^I\, {1\over  U_2^I} \right)\,
\int_0^\infty{dl}\ +\ {\rm finite} . 
\eeqn 
The NSNS tadpole arises from all open string 
sectors in the annulus 
and will be dealt with in section \ref{ch4}, when the scalar 
potential is discussed. \\ 

Since  in type 0$'$ string theory there exists one RR-form of each even degree,
in the original T-dual picture with magnetic fluxes on the 
D9- and D9$'$-branes we have to cancel
the D9- as well as the effective D7-, D5- and D3-brane charges, 
leading to eight separate tadpole cancellation conditions. 
Note, that in type I string theory the D7- and D3-brane charges
were automatically canceled  due to the symmetry under $\Omega$. 
Analogously, here the charges of the second set of RR-forms in type 0B
are projected out by $\Omega'$.   
Thus, the cancellation conditions which derive from the above 
RR tadpole computation are very similar to those of type II string theory, 
except for a doubled background D9-brane charge:
\beqn 
\label{tad1}
&& \sum_{a=1}^K{ N_{a} \prod_{I=1}^3{ n_a^I } } = 
32, \quad 
\sum_{a=1}^K{ N_{a} \prod_{I=1}^3{ \om_a^I } } = 
0, \non
&& \sum_{a=1}^K{ N_{a} n_a^I n_a^J \om_a^K } = 
0, \quad 
\sum_{a=1}^K{ N_{a} n_a^I \om_a^J \om_a^K } = 
0 . 
\eeqn
It is understood that the conditions are to be satisfied for any combination 
of indices 
$I\not= J\not= K\not= I$. The gauge group 
\beqn
G = \prod_{a=1}^K{ U(N_a) } 
\eeqn
may have been reduced in its rank, where 
rk$(G) < 32$ is possible for  $n_a^I > 1$. 
Note, that in contrast to intersecting brane world models of type I string
theory one cannot obtain $SO(2N_a)$ or $Sp(N_a)$ gauge groups. \\ 

\section{Chiral fermion spectra and scalar potential} 
\label{ch4}

The most important feature  for the phenomenological relevance of 
intersecting branes 
is the fact that the open string quantization of strings between two D-branes 
that intersect in a point on the internal space leads to a single chiral 
fermion in the effective lower dimensional theory \cite{hepth9606139}. 

\subsection{Chiral fermions}

This feature gets 
slightly modified in type 0$'$ theory. As was pointed out above, only  open
strings stretched between D6-branes of opposite type give rise
to space-time fermions, whereas open string stretched between two
brane of the same sort give rise to space-time bosons.  
If the two intersecting 
branes are not identified by ${\cal R}$, i.e. at an $(ab')$ or $(a'b)$ 
intersection, 
the fermion will simply transform in the bifundamental representation of the 
respective gauge groups, e.g. the $({\bf N}_{a}, {\bf N}_{b'})$ of $U(N_{a}) 
\times U(N_{b'})$. 
If we are facing an $(aa')$ intersection instead, 
a further distinction has to be made. 
For intersection points that are fixed under ${\cal R}$, one needs to 
regard the projection by $\Omega'{\cal R}$. 
The M\"obius strip amplitude (\ref{ms}) implies the solution
\beqn
\gamma_{\Omega'{\cal R}} = 
\bigotimes_{a=1}^K 
\left( \begin{array}{c c} 
0 & {{\bf 1}}_{N_{a}} \\ {\bf{1}}_{N_{a'}} & 0 
\end{array} \right) 
\eeqn
for the action of $\Omega'{\cal R}$ on the Chan-Paton labels. This leads 
to a single fermion in the antisymmetric representation of $U(N_{a})$ at 
such an 
invariant intersection of type $(aa')$. On the contrary, for intersections not 
invariant under ${\cal R}$ no projection applies and they provide a symmetric 
and 
an antisymmetric representation for any doublet of intersection points.
After defining 
\beqn
I_{ab} = \prod_{I=1}^3{ \left( n_b^I \om_a^I - n_a^I \om_b^I  \right) } , 
\quad 
I^{({\cal R})}_{aa'} = \prod_{I=1}^3{ (2 \om_a^I)} 
\eeqn
for the total and ${\cal R}$-invariant intersection numbers of branes 
$a$ and $b$ or $a'$, 
the spectrum of chiral fermions is summarized in table \ref{spec}. \\

\begin{table}[h]
\begin{center} 
\begin{tabular}{|c|c|}
\hline
& \\[-.40cm]
Representation & Multiplicity \\[.1cm]
\hline
\hline 
& \\[-.3cm]
$\left( {\bf A}_a \right)_L$ & $ I^{({\cal R})}_{aa'}$ \\[.25cm]  
$\left( {\bf A}_a \oplus {\bf S}_a\right)_L$ & 
$ \frac{1}{2} \left( I_{aa'} - I_{aa'}^{({\cal R})} \right)$ 
\\[.25cm] 
$\left( {\bf N}_a,{\bf N}_b \right)_L$ &  $ I_{ab'} $ \\[.25cm]   
\hline 
\end{tabular}
\end{center}
\caption{Chiral massless fermions}
\label{spec}
\end{table}

In contrast to the intersecting brane world models of type I strings, 
where it was necessary to have $b^I > 0$ for at least one torus in order 
to achieve odd numbers of generations, one can now also get three 
generation models with purely imaginary complex structures. 
It is interesting to note that the charged scalar fields can transform in 
different 
representations than the fermions. They arise in the sector of open strings 
with both ends on the same kind of brane and never experience the projection 
by $\Omega'{\cal R}$. 
Thus, space-time bosons and in particular the Higgs fields can  never 
appear in antisymmetric or symmetric representations 
of the gauge group. \\ 

\subsection{Anomaly cancellation}

It is not difficult to check that the spectrum of table \ref{spec} satisfies 
the cancellation of non-abelian anomalies. By using (\ref{tad1}) 
the sum of all contributions to the triangle anomaly of the factor $U(N_a)$ 
vanishes: 
\beqn
\sum_{b'\not= a'}{N_{b'}\, I_{ab'}} + (N_a -4)\, I_{aa'}^{({\cal R})} 
+ 2 N_a\, \frac{1}{2}\, \left( I_{aa'} - I_{aa'}^{({\cal R})}\right) =0 . 
\eeqn
This may serve as a consistency check, but is, of course, guaranteed by the 
tadpole cancellation anyway. \\ 

As usual, the mixed $U(1)-SU(N)^2$ and $U(1)^3$ anomalies 
do not cancel right away but require a suitable Green-Schwarz
mechanism. For the $U(1)_a-SU(N_b)^2$ anomaly one obtains
\beqn \label{ano}
A_{aa}&=&{1\over 2} 
\sum_{b'\not= a'}{N_{b'}\, I_{ab'}} + {(N_a -2)\over 2}\, 2\, 
       I_{aa'}^{({\cal R})} 
+ 2 N_a\, \frac{1}{2} \left( I_{aa'} - I_{aa'}^{({\cal R})}\right) , \non
A_{ab}&=& {1\over 2} N_{a} I_{ab'} \quad {\rm for}\ b\ne a.
\eeqn
Using the relation (\ref{ano}) this can be written as
\beqn \label{anob}
A_{ab}= {1\over 2} N_{a}\, I_{ab'} \quad {\rm for\ all}\ a,b.
\eeqn
Analogous to the type I string, 
we expect that this anomaly is canceled by a generalized Green-Schwarz
mechanism invoking the coupling of four-dimensional bulk RR-fields.
This issue is more conveniently addressed in the original type 0$'$ 
theory, before any T-duality. The spectrum of RR-forms has been given 
in (\ref{actc}). 
In detail, in type 0$'$ string theory one has the following Wess-Zumino terms
in the Born-Infeld action
\beqn \label{anoc}
&&\int_{{\rm D}9_a}  C_0^-\, F_a^5, \quad  
\int_{{\rm D}9_a}  C_2^+\, F_a^4,\quad 
\int_{{\rm D}9_a} C_4^-\, F_a^3, \non
&&\int_{{\rm D}9_a}  C_6^+\, F_a^2, \quad  
\int_{{\rm D}9_a}  C_8^-\, F_a,\quad \int_{{\rm D}9_a} C_{10}^+ .
\eeqn
These forms give rise via dimensional reduction to the following
two-forms and axionic scalars in four dimensions
\beqn \label{anod}
&&C_2^I=  \int_{\mathbb{T}^2_I} C_4^- , \quad\quad  
C_0^I = \int_{\mathbb{T}^2_I} C_2^+ , \non
&&B_2^I=  \int_{\mathbb{T}^2_J\times \mathbb{T}^2_K} C_6^+ ,
\quad\quad  B_0^I=  
\int_{\mathbb{T}^2_J\times \mathbb{T}^2_K} C_4^- , \non
&&B_2=  \int_{\mathbb{T}^6} C_8^- ,\quad\quad  
B_0=  \int_{\mathbb{T}^6} C_6^+ 
\eeqn
and $C_2 = C_2^+,\ C_0 = C_0^-$ 
with the following Hodge-duality relations in four dimensions 
\beqn \label{anoe}
dC_0=\star dB_2, \quad dB_0^I=\star dC^I_2, \quad dC_0^I=\star dB^I_2, \quad
dB_0=\star dC_2. 
\eeqn
Thus, by integrating (\ref{anoc}) over the internal
six-dimensional torus, one obtains the following couplings
\beqn \label{anof}
&&N_a \om_a^1 \om_a^2 \om_a^3 \int_{M_4}  C_2 F_a, \quad
  n_b^1 n_b^2 n_b^3 \int_{M_4}   B_0 F_b^2, \non
&&N_a n_a^I \om_a^J \om_a^K \int_{M_4}  C_2^I F_a, \quad
  n_b^J n_b^K \om_b^I \int_{M_4}  B^I_0 F_b^2, \non
&&N_a n_a^J n_a^K \om_a^I  \int_{M_4}  B_2^I F_a, \quad
  n_b^I \om_b^J \om_b^K \int_{M_4}  C^I_0 F_b^2, \non
&&N_a n_a^1 n_a^2 n_a^3 \int_{M_4}   B_2 F_a, \quad
  \om_b^1 \om_b^2 \om_b^3 \int_{M_4}  C_0 F_b^2 , 
\eeqn
which combine into the usual tree diagrams to contribute 
to the respective anomaly. 
Adding up all terms $F_a\, F_b^2$ we finds that the resulting GS amplitude 
is proportional to 
\beqn \label{anog}
A_{GS}= N_a I_{ab'}
\eeqn
which has the correct form to cancel the field theory anomaly (\ref{anob}).

\subsection{The disc level scalar potential}

The scalar potential can be read off from the dilaton tadpole divergence of 
the closed 
string tree channel amplitude which is essentially the square of the disc 
expectation value 
\beqn
\langle \phi \rangle_{\rm disc} \sim \frac{\partial V(\phi,U_2^I)}{\partial 
\phi}. 
\eeqn
Due to the absence of the NSNS contributions 
in the Klein bottle and M\"obius strip diagrams, the resulting potential is 
identical to the one obtained in type I string theory \cite{hepth0107138} 
except for the absence of the orientifold tension, 
\beqn \label{pot}
V(\phi, U_2^I) = T_6 e^{-\phi_4} \sum_{a=1}^K{ (2N_a) 
\prod_{I=1}^3{\sqrt{ \left(n_a^I\right)^2 U_2^I + 
\frac{( \om_a^I )^2}{U_2^I} } }
} ,  
\eeqn
where $\phi_4$ is the four-dimensional dilaton 
$\phi_4= \phi-\ln ({\rm Vol}_6)/2$ and $T_6$ the 
D6-brane tension. The normalization was fixed by comparing to the DBI 
effective action.  
Already from the scaling behavior 
of the potential 
\beqn
V(\phi, U_2^I) \longrightarrow \infty \quad {\rm for}\quad U_2^I 
\longrightarrow 0,\infty 
\eeqn  
it is clear that there must exist a global minimum at which the $U_2^I$ are 
stabilized at tree level. We will demonstrate this explicitly by presenting an 
example with stabilized complex structure in the following section. \\ 

The potential (\ref{pot}) does not depend on the K\"ahler moduli $T^I$. 
It was explained in \cite{hepth0107138} that it is to be expected that this 
feature remains true perturbatively if the D6-branes 
intersect in points. Only if there are also branes that are parallel on 
circles of the 
internal space the propagation of KK and winding modes will introduce a 
dependence 
on the $T^I$. It was estimated from their respective proportionality to 
$1/T^I$ and  
$T^I$ that at least to the next to leading order, in the annulus diagram, 
the K\"ahler moduli may stabilize as well. 

\section{A left-right symmetrically unified model}
\label{ch5}

In this section we present a concrete model which contains the gauge group
of a left-right symmetric extension of the Standard Model with some 
additional charged chiral matter.
Since, in contrast to the type II orientifold models, the type 0$'$ orientifold
models only have chiral fermions for $ab'$ intersections, the anomaly 
cancellation
conditions for the $U(N_a)$ gauge factors (including $N_a=1,2$) prevent 
the realization of a pure three generation Standard Model. Recall
that for the type II orientifold models two of the left-handed quark doublets
transformed in the $(\bf 3,2)$ representation and one in the 
$(\bf \overline{3},2)$ representation. In this way, the $(ab)$ and $(ab')$ 
intersections could combine to provide three generations, which is 
now impossible, as only $(ab')$ will provide fermions at all. \\ 

We now choose five stacks of D6-D6$'$ branes, $b^I=1/2$ on 
all three two-dimensional
tori $\mathbb{T}^2_I$ and the wrapping numbers shown in table 2.

\begin{table}[h]
\label{tab2}
\begin{center} 
\begin{tabular}{|c|c|c|c|}
\hline 
&&& \\[-.40cm]
$(n_a^1,\om_a^1)$ & $(n^2_a,\om^2_a)$ & $(n_a^3,\om^3_a)$   & $N_a$ \\[.1cm]
\hline
\hline 
&&& \\[-.4cm]
$(2,0)$ & $(1,-{1\over 2})$ & $(4,1)$ &   $3$ \\[.1cm]
$(1,{1\over 2})$ & $(1,{3\over 2})$ & $(1,{1\over 2})$ &   $2$ \\[.1cm]
$(1,-{1\over 2})$ & $(1,{3\over 2})$ & $(1,{1\over 2})$ &   $2$ \\[.1cm]
$(2,0)$ & $(1,{7\over 2})$ & $(1,-{1\over 2})$ &   $1$ \\[.1cm]
$(2,0)$ & $(1,-{1\over 2})$ & $(1,-{7\over 2})$ &   $1$ \\[.1cm]
\hline
\end{tabular}
\end{center}
\caption{Wrapping numbers}
\end{table}

It can easily be checked that these wrapping numbers satisfy the RR-tadpole
cancellation conditions. The resulting massless chiral spectrum is
presented in table 3. \\

\begin{table}[h]
\label{tab3}
\begin{center} 
\begin{tabular}{|l|c|}
\hline 
& \\[-.40cm]
$SU(3)\times SU(2)_L\times SU(2)_R\times U(1)^5$  & Multiplicity \\[.1cm]
\hline
\hline 
& \\[-.4cm]
$\quad\quad ({\bf 3},{\bf 2},{\bf 1})_{(1,1,0,0,0)}$ &   $3$ \\[.1cm]
$\quad\quad ({\bf \overline{3}},{\bf 1},{\bf 2})_{(-1,0,-1,0,0)}$ &  $3$ \\[.1cm]
\hline
& \\[-.4cm]
$\quad\quad ({\bf 1},{\bf 2},{\bf 1})_{(0,-1,0,0,-1)}$ &  $3$ \\[.1cm]
$\quad\quad ({\bf 1},{\bf 1},{\bf 2})_{(0,0,1,0,1)}$ &  $3$ \\[.1cm]
\hline
& \\[-.4cm]
$\quad\quad ({\bf 1},{\bf A},{\bf 1})_{(0,2,0,0,0)}$ &  $3$ \\[.1cm]
\hline
& \\[-.4cm]
$\quad\quad ({\bf 1},{\bf 1},{\bf \overline{A}})_{(0,0,-2,0,0)}$ &  $3$ \\[.1cm]
\hline
\end{tabular}
\end{center}
\caption{Chiral massless spectrum}
\end{table}

This is the matter content of the minimal left-right symmetric extension of 
the Standard Model plus three generations of a charged 
$({\bf A,1}) \oplus ({\bf 1,\overline{A}})$ of $SU(2)_L \times SU(2)_R$. 
Note, that the fourth $U(1)$ factor decouples completely from the chiral
spectrum and therefore can be considered as resulting from a spectator
brane which is only there to satisfy the RR cancellation conditions.
Two of the five $U(1)$ factors are anomalous and besides $U(1)_4$ the  
anomaly free ones are
\beqn
 U(1)_{B-L}&=&{1\over 3}\, U(1)_1 + U(1)_5 , \non
 U(1)_K&=&U(1)_2 +U(1)_3+2\, U(1)_5.
\eeqn 
It is beyond the scope of this paper to follow the phenomenological properties
of this model further. Instead, we investigate the
disc level scalar potential for this concrete model to provide an 
explicit example for the stabilization of the complex structure moduli. \\ 

By performing a numerical analysis of the resulting potential (\ref{pot})
we find that there exist a unique global minimum for the values
of the complex structures
\beqn
U_2^1=0.086,\quad U_2^2=1.045,\quad U_2^3=0.486.
\eeqn 

We have also determined the ground state energies $E_0$ in the various 
bosonic open string sectors at this minimum 
where depending on the values of the complex structures tachyons can 
potentially arise.
It is another nice feature of these type 0$'$ models that only $(ab)$ type 
intersections need to be considered, thus the number of dangerous sectors 
is essentially halved. 
If two D6-branes are parallel on at least one $\mathbb{T}^2$, then 
we can move the two D-branes apart on this torus
and avoid the tachyonic instability classically.
For the remaining open string sectors we find the
following ground state energies: \\ 

\begin{table}[h]
\begin{center} 
\begin{tabular}{|c|c|}
\hline 
& \\[-.4cm]
Sector   & $E_0$ \\
\hline
\hline 
& \\[-.4cm]
$(12)$ &   $0.05$ \\[.1cm]
$(13)$ &  $0.05$ \\[.1cm]
$(24)$ &   $0.02$ \\[.1cm]
$(34)$ &  $0.02$ \\[.1cm]
$(25)$ &   $0.09$ \\[.1cm]
$(35)$ &  $0.09$ \\[.1cm]
\hline
\end{tabular}
\end{center}
\caption{Ground state energies}
\end{table}

All ground state energies are positive implying that this model, 
in the approximation we used, is perturbatively stable.
Note, that the possible tachyon in the $(23)$ open string sector
transforms in the $(\bf 2,\bf 2)$ representation and is needed for
breaking the left-right symmetric down to the Standard Model.
However, as we mentioned already there can be no tachyons and therefore
no Higgs particles in symmetric or anti-symmetric representations
of the gauge group. \\

\section{Conclusions}

In this paper we have constructed a class of non-supersymmetric string 
models whose geometric complex structure 
moduli are stabilized at finite values by the leading order 
perturbative potential. One may further argue, that the stabilization 
can be extended to the K\"ahler moduli as well, hence leaving only the dilaton 
unstable perturbatively. 
Our novel construction is intrinsically non-supersymmetric, as it starts 
from the non-supersymmetric but tachyon-free ten-dimensional type 0$'$ 
string theory compactified on a torus. 
To this theory we have applied the techniques of intersecting 
brane worlds finding a set of solutions for the effective four-dimensional 
field theory which displays perspectives for obtaining semi-realistic 
models that are comparative to the earlier studied type I and type II 
compactifications. 
One may consider various extensions and generalizations of the present 
program. It may for instance be interesting to construct 
orbifolds of the purely toroidal 
models along the lines of \cite{hepth9912204,hepth0008250}.

\vspace{1cm}
\begin{center}
{\bf Acknowledgements} \\
\end{center}
We are supported in part by the EEC RTN programme HPRN-CT-2000-00131. 
We would like to thank A. Uranga for discussion.\\ 

\begin{appendix}

\section{Amplitudes of type 0$'$ with branes at angles}

In this appendix we summarize the more technical results for the three 
amplitudes 
that contribute to the tadpole divergence at the $\chi =0$ level of 
perturbation theory. 
They are obtained most directly by combining the results of 
\cite{hepth9904069} and \cite{hepth0007024}. 
The Klein bottle amplitude of the original ten-dimensional type 0$'$ theory 
was 
given in (\ref{kb}). The modification due to the toroidal compactification 
is standard and does not take reference to the novel issue of intersecting 
branes,
\beqn 
\widetilde{\cal K} = -2^7 c_4\, \prod_{I=1}^3{\left( U_2^I \right)}\ 
\int_0^\infty{dl\ 
\frac{\vartheta [ {1/2 \atop 0} ]^4}{\eta^{12}} 
\prod_{I=1}^3{\left( \sum_{r,s\in \mathbb{Z}}{ 
               e^{-4\pi l\left( r^2 {R_1^I}^2 + s^2/ {R_2^I}^2 \right) } } 
\right) }} . 
\eeqn
For the open string amplitude the peculiarities of type 0$'$ as well as the 
modifications due to the relative angles of the D6-branes need to be taken 
into 
account. The open string states involve only the NS sector but one needs to 
include the D6- and D6$'$-branes now. 
The relative angles lead to shifts in the oscillator spectrum 
of string excitations by 
\beqn
\epsilon_{ab}^I = \frac{\varphi_b^I - \varphi_a^I}{\pi}
\eeqn
and change the KK and winding zero mode quantization 
\cite{hepth0003024}. The annulus diagram for strings with both ends on the 
same brane $a$ then reads
\beqn \nonumber
\widetilde{\cal A}_{aa} = 2^{-4}\,c_4\, N_a^2
\prod_{I=1}^3{ \left( ( n_a^I)^2 U_2^I + 
                    \frac{( \om_a^I)^2}{U_2^I} \right) }  
\int_0^\infty{dl\  \frac{\vartheta [ {0 \atop 0} ]^4 - 
\vartheta [ {1/2 \atop 0} ]^4}
                    {\eta^{12}} 
               \prod_{I=1}^3{\left( 
        \sum_{r,s\in \mathbb{Z}}{ e^{-\pi l \widetilde{M}_{I}^{2}}}\right) } } 
\eeqn
which is to be combined with an identical contribution from the $a'$ branes. 
The annulus diagram for strings between two different branes 
$a$ and $b$ (or $a'$ and $b'$) but both of the same type is
\beqn
\widetilde{\cal A}_{ab} =  2^{-1}\, c_4\, N_a N_b I_{ab} 
\int _0^\infty{dl\ \frac{\vartheta [ {0 \atop 0} ] 
                         \prod_I{\vartheta [ {0 \atop \epsilon_{ab}^I} ]} -
                         \vartheta [ {1/2 \atop 0} ] 
                         \prod_I{\vartheta [ {1/2 \atop \epsilon_{ab}^I} ]}}{
           \eta^3\prod_I{\vartheta [ {1/2 \atop 1/2 + \epsilon_{ab}^I} ]} }}, 
\eeqn
while for open strings between branes $a$ and $b'$ (or $a'$ and $b$) one finds
\beqn
\widetilde{\cal A}_{ab'} = - 2^{-1}\, c_4\, N_a N_{b'} I_{ab'} 
\int _0^\infty{dl\ \frac{\vartheta [ {0 \atop 1/2} ] 
                   \prod_I{\vartheta [ {0 \atop 1/2 + \epsilon_{ab'}^I} ]}}{
          \eta^3\prod_I{\vartheta [ {1/2 \atop 1/2 + \epsilon_{ab'}^I} ]} }} .
\eeqn
Together this combines into the annulus diagram of type II 
strings, which in particular guarantees the absence of any tachyon 
contribution to the tadpole divergence. 
Finally, the M\"obius strip diagram gives the contribution 
\beqn 
\widetilde{\cal M}_{aa'} = 2^2 c_4 N_a I_{aa'}^{({\cal R})} 
\int_0^\infty{dl\ \frac{\vartheta [ {1/2 \atop 0} ] 
                   \prod_I{\vartheta [ {1/2 \atop \varphi_a^I/ \pi} ]}}{
          \eta^3\prod_I{\vartheta [ {1/2 \atop 1/2 + \varphi_a^I/ \pi} ]} }} .
\eeqn
We have used the standard definitions 
\beqn
\frac{\vartheta[ {\alpha \atop \beta} ] (q)}{\eta (q)} 
&=& e^{2\pi i\alpha\beta} q^{{\alpha^2\over 2}-1/24} 
\prod_{n=1}^\infty{ \left( \left( 1+q^{n-1/2+\alpha} e^{2\pi i\beta} \right) 
                           \left( 1+q^{n-1/2-\alpha} e^{-2\pi i\beta} \right)
\right) } , \non
\eta (q) &=& q^{1/24} \prod_{n=1}^\infty{\left( 1-q^n \right)} . 
\eeqn 
                         
\end{appendix}

\bibliography{articles}
\bibliographystyle{unsrt}

\end{document}